\documentclass[12pt]{article}
\usepackage{mathrsfs}

\usepackage{graphicx}

\begin{document}

\title{Dyson-Schwinger Equation and Quantum Phase Transitions in Massless QCD  }

\author{{Wei Yuan$^{a}$, Huan Chen$^{a}$,
and Yu-xin Liu$^{a,b,c,}$\thanks{Corresponding author, e-mail address: yxliu@pku.edu.cn} }\\[3mm]
\normalsize{$^1$ Department of Physics, Peking University, Beijing 100871, China}\\
\normalsize{$^2$ The Key Laboratory of Heavy Ion Physics,
Ministry of Education, Beijing 100871, China } \\
\normalsize{$^3$ Center of Theoretical Nuclear Physics, National
Laboratory of Heavy Ion Accelerator,}\\ \normalsize{ Lanzhou 730000,
China}  }

\maketitle

%\author{Wei Yuan}
%\affiliation{Department of Physics, Peking University, Beijing
%100871, China}
%
%\author{Huan Chen}
%\affiliation{Department of Physics, Peking University, Beijing
%100871, China}
%
%\author{Yu-xin Liu}
%\email[corresponding author]{} \affiliation{Department of Physics,
%Peking University, Beijing 100871, China}
%\affiliation{The Key Laboratory of Heavy Ion Physics,
%Ministry of Education,Beijing 100871, China }
%\affiliation{Center of Theoretical Nuclear Physics, National
%Laboratory of Heavy Ion Accelerator, Lanzhou 730000, China}

%\date{\today}

\begin{abstract}
We study the stability of the highest symmetric solution
(Wigner-solution) of Dyson-Schwinger equations in chiral limit and
at zero temperature.  Our results confirm that if the chemical
potential is not very large, the QCD vacuum is in the chiral
symmetry breaking phase and the quantum phase-transition of the
chiral symmetry restoration is in first order. Meanwhile it seems
that there is not competition between chiral symmetry breaking phase
and color superconductivity phase since the color superconductivity
phase appears only if the chemical potential is very large.
Moreover, we propose that chiral symmetry breaking arises from the
positive feedback with respect to the mass perturbation.
\end{abstract}

%\pacs{12.38.Lg, 11.30.Rd, 11.10.Wx, 25.75.Nq}


%\maketitle

\newpage

\parindent=20pt
It has been known that the results of perturbative
renormalization-group in QCD will encounter divergence at low
energy region. This behavior indicates an important fact that the
vacuum (Wigner-vacuum) defined within Feynman's original
path-integral framework which possesses the highest symmetries is
unstable and incorrect\cite{Wilzeck}. In addition, it should be
emphasized that the effective action (viz. minus effective
potential) based on Wigner-vacuum should have the same globle and
modified BRST symmetry\cite{BRST} as the original
action\cite{symmetry}. Thus, the gaps corresponding to the running
coupling coefficients of the variational one-particle-irreducible
(1PI) vertexes, which break the global symmetry and BRST gauge
symmetry, should definitely be zero! Nevertheless, the statement,
which is correct in high energy region and is the key to prove
gauge theory is perturbatively renormalizable\cite{BRST}, often
deviates from observations to real QCD in low energy region.
Therefore, the real QCD vacuum in low energy channel is different
from the Wigner-vacuum. Anyway, we should note that energy or free
energy (if chemical potential $\mu\neq 0$) of Wigner-vacuum is
definitely the lowest in Euclidean formalism of QCD at zero
temperature (temperature $T=0$ means a unbounded imaginal time).
Thus we can't justify a phase-transition by comparing the energy
between the symmetrical vacuum and the symmetry-broken one.
Indeed, there has been a powerful general argument that, while
ground states of a macroscopic system are degenerate, the true
vacuum is the asymmetric one rather than the Wigner-vacuum because
the Wigner-vacuum is extremely unstable against a perturbation
(viz. symmetric system will undergo a dramatic transforms against
an even arbitrarily small perturbation)\cite{stable}. However,
such an idea has not yet been proved solidly. It is then
interesting to study under what conditions the vacuums of QCD tend
to be degenerate, and how to understand the phase-transitions in a
reasonable and systematic way. It has been known that
Dyson-Schwinger (D-S) equations approach provides a
nonperturbative framework to study the vacuum properties of strong
interaction and hadron properties in free space and to simulate
the chiral symmetry restoration and deconfinement in the system
with finite temperature and/or finite chemical
potential\cite{Roberts00,Roberts94,Alkofer01,Roberts03}. In this
paper, we intend to shed some light on these questions by
analyzing the stability of Wigner-vacuum in the framework of
Dyson-Schwinger equation approach.

Within Feynman's framework, one can easily derive a series of
dynamical integral-equations (Dyson-Schwinger
equations)\cite{Roberts94} and a series of identities which come
from the symmetries. If we combine these two series of results, we
will get a theory based on Wigner-vacuum which gives Wigner solution
of the equations for any correlations and gaps. On the other hand,
if we relax the restrictions of the symmetries, the Dyson-Schwinger
(D-S) equation will be correct not only for the symmetrical vacuum
but also for the asymmetrical vacuum (if existing!) and, besides the
Winger solution, we can get a new class of solutions (Nambu
solution) of the D-S equation for the correlations and gaps. We must
note here that the Nambu solution is important if and only if the
Wigner-vacuum is unstable under perturbations, viz. vacuums are
degenerate, otherwise, Nambu solution is not the real ground state
but a dynamically stable excited state. For example, as we will see
at below, the fact that chiral susceptibility of Wigner mass
function in the chiral limit is negative indicates the
chiral-symmetry breaking of the QCD vacuum in a natural sense, and
the restoration of chiral symmetry is closely related to the
presence of positive chiral susceptibility beyond some critical
chemical potential.

It is well known that D-S equation is a series unclosed equations
where the equation for $n$-point Green functions depend on $(n+1)$
point Green functions. One should then truncate it with
indispensable approximations before taking it to evaluate any
physics quantity practically. Furthermore, in order to obtain a full
and reliable understanding of QCD phase-diagram with respect to the
medium density (or the chemical potential), we should investigate
the stability of the Wigner-vacuum against all possible
perturbations simultaneously which are allowed by the truncation to
D-S equation. In this work, for simplicity, we take the rainbow
approximation and investigate the stability of the Wigner-vacuum
against the allowed chiral-perturbation and diquark-perturbation.

We begin with single-flavor QCD in chiral limit. The action in
Euclidean space is usually given as
\begin{equation}
\label{action}
 S=\int d^{4} x[\bar{\Psi}i\gamma\cdot D
  \Psi+\frac{1}{4}F^{a}_{\mu\nu}F^{a}_{\mu\nu}-i\mu\bar{\Psi}\gamma_{0}\Psi]
\end{equation}
where the $\gamma$ matrixes are chosen to satisfy
$[\gamma_{\mu},\gamma_{\nu}]_{+}=-2\delta_{\mu\nu}$
and $D_{\mu}=\partial_{\mu}+ig_{\Lambda}\frac{\lambda_{a}}{2}A_{a\mu}$.

Representing the 1PI vertexes of $\bar{\psi}\psi$, $\psi\psi$,
$\bar{\psi}\bar{\psi}$ as $\Sigma_{1}$, $\Sigma_{2}$,
$\Sigma_{3}$, respectively, we can write the dressed quark
propagator without diquark component as
\begin{equation}
\tilde{G}=\frac{1}{p\cdot \gamma-i\mu\gamma_{0}+\Sigma_{1}}.
\end{equation}
Meanwhile the relations between the 1PI vertexes and the full propagators
can be written as
\begin{equation}
G_{1}=\frac{\tilde{G}}{1-\Sigma_{3}\tilde{G}\Sigma_{2}\tilde{G}},
\quad G_{2,3}=\tilde{G}(-\Sigma_{2,3})G_{1}.
\end{equation}
With these relations we can easily obtain the corresponding D-S equation
\begin{equation}
\Sigma_{1,2,3}(p)=-g_{\Lambda}^{2}\int\frac{d^{4}q}{(2\pi)^4}\gamma_{\mu}
\frac{\lambda_{\eta}}{2}G_{1,2,3}(q)\gamma_{\nu}\frac{\lambda_{\eta}}{2}
D_{\mu\nu}(k) \, ,
\end{equation}
where $k \equiv p-q$ stands for the momentum transferred.
Furthermore, if we are only interested in the condensates in chiral
and color $\bar{3}$ diqurak channels which preserves the parity and
in which single gluon exchange interaction is
attractive\cite{Wilzeck}, the 1PI vertexes in strong interaction
matter with chemical potential $\mu$ can be expressed as
\begin{equation}
\Sigma_{1}(p)=[A(p)-1]\vec{p}\cdot\vec{\gamma} +
[C(p)-1](p_{0}-i\mu)\gamma_{0} + \Delta_{1}(p) \, ,
\end{equation}
\begin{equation}
\Sigma_{2}(p)=\Delta_{2}(p) M\gamma_{5}\hat{C} \, ,
\end{equation}
\begin{equation}
\Sigma_{3}(p)=\Delta_{3}(p) \hat{C} M\gamma_{5}\, ,
\end{equation}
where $M$ is a matrix in color space corresponding to the color
$\bar{3}$ diquark channel, $\hat{C}$ is the charge conjugation
operator, which can be given explicitly as
\begin{equation}
M_{\alpha\beta}=\epsilon_{1\alpha\beta}
\end{equation}
\begin{equation}
\hat{C}\Psi=\gamma_{2}\gamma_{0}\bar{\Psi}^{T} \, , \qquad
\hat{C}\bar{\Psi}=\Psi^{T}\gamma_{2}\gamma_{0} \, .
\end{equation}
It is evident that the Wigner solution is characterized by
$\Delta_{1,2,3} \equiv 0$.

Instead of exploiting Nambu solution of D-S equation, we prefer to
investigate whether the Wigner-vacuum is stable against
perturbations of the condensates $\Delta^{b}_{1}\bar{\Psi}\Psi$,
$\Delta^{b}_{2}\bar{\Psi}M \gamma_{5}\hat{C}\Psi$ and
$\Delta^{b}_{3}(\hat{C}\bar{\Psi}) M\gamma_{5}\Psi$. For our
purpose, here we have defined the perturbed Wigner solutions for the
gaps as
\begin{equation}
\Delta_{1}^{Wigner}(p)=F_{1}(p)\Delta_{1}^{b} \, ,
\end{equation}
\begin{equation}
\Delta_{2}^{Wigner}(p)=F_{2}(p)\Delta_{2}^{b}\, ,
\end{equation}
\begin{equation}\label{}
\Delta_{3}^{Wigner}(p)=F_{3}(p)\Delta_{3}^{b}\, .
\end{equation}
The $F_{1}$, $F_{2,3}$ stands for the susceptibility with respect to
the chiral channel, diquark channels, respectively.

From the D-S equation with infinitesimal but explicit mass term,
we find that, when $\Delta^{b}_{1}\neq 0,\, \Delta_{2,3}^{b}=0$,
$F_{1}$ satisfies equation
\begin{equation}
F_{1}(p)=1-g_{\Lambda}^{2}\int\frac{d^{4}q}{(2\pi)^4}\gamma_{\mu}
\frac{\lambda_{\eta}}{2}\frac{F_{1}(q)}{A_{W}^{2}(q)\vec{q}^{2}+
C_{W}^{2}(q)(q_{0}-i\mu)^{2}}\gamma_{\nu}
\frac{\lambda_{\eta}}{2}D_{\mu\nu}(k) \, .
\end{equation}
From the D-S equation with infinitesimal but explicit {\it
diquark} term, we obtain that, when $\Delta^{b}_{2}\neq 0, \;
\Delta_{1,3}^{b}=0$, $F_{2}$ satisfies equation
\begin{eqnarray}
F_{2}(p)M\gamma_{5}\hat{C}&=&M\gamma_{5}\hat{C}+g_{\Lambda}^{2}
\int\frac{d^{4}q}{(2\pi)^4}\gamma_{\mu}\frac{\lambda_{\eta}}{2}
\frac{F_{2}(q)}{A_{W}(q)\vec{q} \cdot \vec{\gamma}
+ C_{W}(q)(q_{0}-i\mu)\gamma_{0}}  \nonumber \\
& & \times M\gamma_{5}\hat{C} \frac{1}{A_{W}(q)\vec{q}\cdot
\vec{\gamma}+ C_{W}(q)(q_{0}-i\mu)\gamma_{0}}\gamma_{\nu}
\frac{\lambda_{\eta}}{2}D_{\mu\nu}(k) \, .
\end{eqnarray}
Here $A_{W}$, $C_{W}$ are Wigner solutions for the $A$, $C$ which
satisfy the D-S equation with $\Delta _{1,2,3} \equiv 0 $. After
some derivation, the equations can be explicitly written as
\begin{equation}
\label{WA} A_{W}(p)=1-\frac{4g_{\Lambda}^{2}}{3\vec{p}^{2}} \! \int \!\!
\frac{d^{4}q}{(2\pi)^4} \frac{A_{W}(q)[2p_{i}q_{j}D_{ij}(k)
-\vec{p}\cdot\vec{q}D_{\mu\mu}(k)]+2C_{W}(q)
[(q_{0}-i\mu)p_{i}D_{i0}(k)]}{A_{W}^{2}(q)\vec{q}^{2}
+C_{W}^{2}(q)(q_{0}-i\mu)^{2}} \, ,
\end{equation}
and
\begin{equation}
\label{WB} C_{W}(p)=1-\frac{4g_{\Lambda}^{2}}{3(p_{0}\! - \! i\mu)^{2}}
\! \int \!\!
\frac{d^{4}q}{(2\pi)^4}\frac{2A_{W}(q)(p_{0}\! - \! i\mu)q_{i}D_{0i}(k)
+C_{W}(q)(p_{0}\! - \! i\mu)(q_{0}\! - \! i\mu)D_{\mu\mu}^{M}(k)}
{A_{W}^{2}(q)\vec{q}^{2}+C_{W}^{2}(q)(q_{0}-i\mu)^{2}} \, ,
\end{equation}
where $D^{M}_{\mu\mu}\equiv D_{00}-D_{ii}.$

It is apparent that to solve the equation practically, one needs the
effective gluon propagator as an input.
We then take the effective gluon propagator as the T\"{u}bingen model\cite{parameter}
\begin{equation}
D_{\mu\nu}(k)=(\delta_{\mu\nu}-\frac{k_{\mu}k_{\nu}}{k^{2}})D(k) \, ,
\end{equation}
with
$$ g_{\Lambda}^{2}D(k) = 4 \pi ^{2} d \frac{k^{2}}{\omega^{2}}
e^{-\frac{k^{2}}{\omega^{2}} }\, . $$
By using relation
\begin{equation}
\frac{\lambda_{\eta}}{2}M\frac{\lambda_{\eta}^{T}}{2}=-\frac{2}{3}M
\, ,
\end{equation}
we could conclude that the stability of Wigner-vacuum against chiral
and diquark perturbation is characterized by linear integral
equations
\begin{equation}
\label{f1} F_{1}(p)=1+4g_{\Lambda}^{2} \int\frac{d^{4}q}{(2\pi)^4}
\frac{F_{1}(q)D(k)}{A_{W}^{2}(q)\vec{q}^{2} +
C_{W}^{2}(q)(q_{0}-i\mu)^{2}} \, ,
\end{equation}
 and
\begin{equation} \label{f23}
F_{2,3}(p)=1-2g_{\Lambda}^{2} \int\frac{d^{4}q}{(2\pi)^4}
\frac{F_{2,3}(q)D(k)[A_{W}^{2}(q)\vec{q}^{2}+C_{W}^{2}(q)(q_{0}^{2}
+ \mu^{2})]}{[A_{W}^{2}\vec{q}^{2} +
C_{W}^{2}(q_{0}^{2}-\mu^{2})]^{2}+[2C_{W}^{2}\mu q_{0}]^{2}} \, .
\end{equation}

Up to now, what we have discussed is only one flavor of quark. In
fact, we can deal with the case of two or three flavors of quarks
without any technical difficulty in chiral limit. Since in chiral
limit, flavor has nothing to do with dynamics, the gapped flavor
channel should be chosen to leave the maximal unbroken symmetries
\cite{Wilzeck}, such as
 $\bar{\Psi}_{i\alpha}\epsilon^{1\alpha\beta}\epsilon^{ij}
\gamma_{5}\hat{C}\Psi_{j\beta}$ in case of two flavors and
$\bar{\Psi}_{i\alpha}\epsilon_{A}^{\ \
\alpha\beta}\epsilon^{Aij}\gamma_{5}\hat{C}\Psi_{j\beta}$ in case of
three flavors($\alpha,\beta$ are color indexes and $i,j$ are flavor
indexes), which are the well known two-flavor color
superconductivity (2CS) and color flavor locking (CFL) channels,
respectively. However, all these considerations, which do not change
the equations of the susceptibilities in chiral limit, are
nontrivial in case of real mass spectrum of $\{u,d,s\}$ quarks
because it has strongly indicated a transition from the 2CS phase to
the CFL phase at some relatively high chemical potential. In the
present work, as we work in chiral limit, one-flavor should be
sufficient.

To obtain the quantitative result, we solve the D-S equation with
the T\"{u}bingen model of the effective gluon propagator and
parameters $\omega = 0.4$~GeV, $d = 45.0~\mbox{GeV}^{-2}$, with
which the pion properties and some other low energy chiral
observables have been described well\cite{parameter}. The obtained
results of the chemical potential dependence of the Wigner solutions
$A_{W}[\mu]$, $C_{W}[\mu]$ are illustrated in Fig.~1. The obtained
chemical potential dependence of the Nambu dynamical mass function
as the ratio $\Delta_{1N} / A_{N}$ of the Nambu solutions is
illustrated in Fig.~2. And the chemical potential dependence of the
susceptibilities corresponding to the chiral quark gap and diquark
gap of Wigner solution ($F_{1}[\mu]$ and $F_{2}[\mu]$) are displayed
in Fig.~3, Fig.~4, respectively. Fig.~1 shows obviously that the
Wigner-solutions $A(p,\mu)$ and $C(p,\mu)$ are equal only if $\mu=0$
(another point shown in Fig.1 is not universal for other momentum
modes), but remarkably separated from each other once $\mu\neq 0$.
Since the $A_{W}$ and $C_{W}$ contain the crucial information of the
QCD-vacuum in our current framework, we have calculated them as
precisely as we can, rather than taking any approximation such as
$A=C$\cite{Zong05}. In fact, because of the obvious oscillating
behavior of $A_{W}(\mu)$ and $C_{W}(\mu)$ caused by chemical
potential, which will be restrained by chiral gap in Nambu
solutions, the numerical calculation for Wigner solutions is much
more difficult than for Nambu solutions.
\begin{figure}[!htb]
\centering
\includegraphics[scale=1.0]{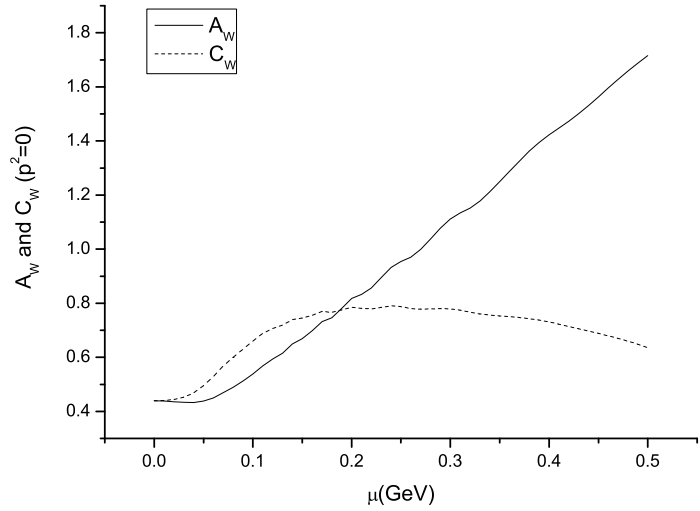}
\caption{\label{1} Chemical potential $\mu$ dependence of
Winger-solutions in extremely low energy channel ($p^{2} = 0$) }
\end{figure}

\begin{figure}[!htb]
\centering
\includegraphics[scale=1.0]{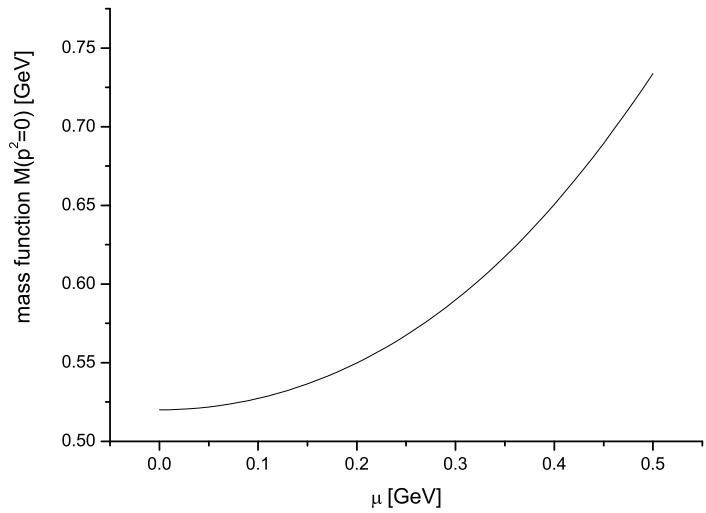}
\caption{\label{2} chemical potential $\mu$ dependence of the
dynamical quark mass function in extremely low energy channel
($p^{2} = 0$)}
\end{figure}

\begin{figure}[!htb]
\centering
\includegraphics[scale=1.0]{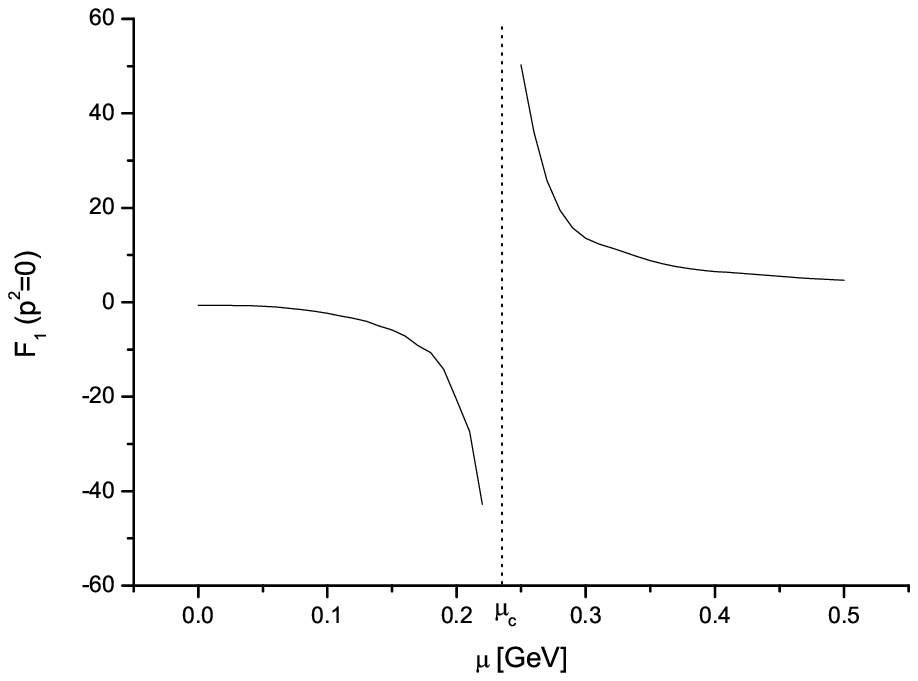}
\caption{\label{3}  Chemical potential $\mu$ dependence of the
susceptibility $F_{1}$ in extremely low energy channel ($p^{2} =
0$)}
\end{figure}

\begin{figure}[!htb]
\centering
\includegraphics[scale=1.0]{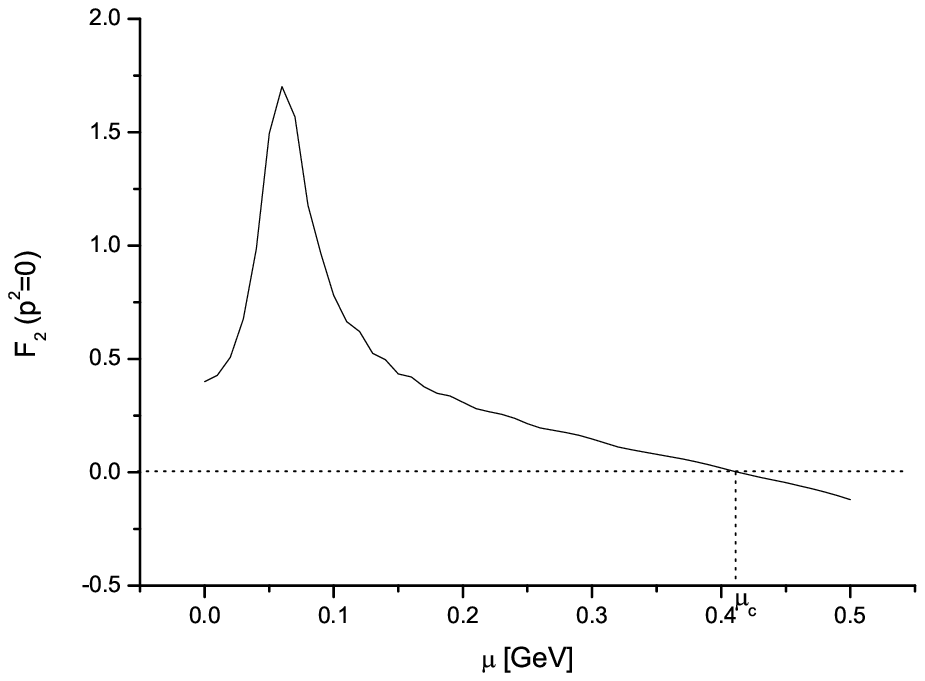}
\caption{\label{4} Chemical potential $\mu$ dependence of the
susceptibility $F_{2}$ in extremely low energy channel ($p^{2} =
0$)}
\end{figure}

Because the susceptibility is the response rate of the dynamical
mass with respect to the perturbation of current quark mass, with
the above obtained result we discuss the quantum phase-transition in
QCD. At first we take, for instance, a free fermion system in chiral
limit with $F_{1}\equiv1$ as a starting point. Corresponding to the
presence of the repulsive potential in chiral channel induced by
mass vertex of the system, the influence of small and positive mass
term on the system is nothing more than a gapped dispersion relation
$E=(\vec{p}^{2}+m^{2})^{1/2}$. Changing the sign of the perturbative
mass term just trivially results in a redefinition of fermion and
anti-fermion, and of course the same physics. Therefore, the
influence of small mass term on free system is really perturbative,
and we can infer safely that the free vacuum should preserve chiral
symmetry. Now, we turn to the case of QCD. Because QCD is an
asymptotically free theory, the chiral susceptibility in high energy
mode is positive to maintain the system stable in high energy
region. If the $F_{1}$ in low energy mode is also {\it positive},
the QCD should be similar to the free case by means of that the
vacuum is stable against mass perturbation. In fact, our calculation
indicates that, in low chemical potential region, the $F_{1}$ is
{\it negative} in low energy channel. In this case, a small and
positive mass term $m\bar{\Psi}\Psi$ results in an attractive (not
repulsive as the free case!) potential in the chiral channel whose
strength is characterized by $|mF_{1}|$. This attractive
interaction, even if quite weak, should induce a small $\langle
\bar{\Psi}\Psi \rangle$ condensate correspondingly. Since the
induced small condensate can be treated as a positive mass-like
vertex, which will be dressed by QCD interaction and results in a
deeper attractive potential (if chiral susceptibility is still
negative) in chiral channel and a ``larger" condensate! Such and
such, the chiral condensate will become ``larger" and ``larger" as
far as the chiral susceptibility become positive. In the limit
$m\rightarrow 0^{+}$, this process tends to be quasistatic, without
change of free energy or entropy or energy, it means that a
spontaneous quantum phase transition from Wigner-vacuum to
Nambu-vacuum takes place! If the perturbative mass is negative, the
situation is similar, but leads the Wigner-vacuum to another
physically equivalent Nambu-vacuum which can be affirmed in the
framework of D-S equation (in the chiral limit, we can always find
two Nambu solutions for mass function with the same quantity and
opposite sign, if the interaction in infrared region is strong
enough\cite{Roberts94,Pennington,CLR05}). From Fig.~3, one can
easily recognize that the susceptibility $F_{1}$ is definitely
negative if the chemical potential $\mu<0.24$~GeV, it becomes
positive if the chemical potential $\mu>0.24$~GeV, and more
significantly, the $F_{1}$ is divergent and disconnected at
$\mu=0.24$~GeV. Meanwhile Fig.~4 indicates that the susceptibility
$F_{2}$ is positive if the chemical potential $\mu < 0.42$~GeV and
it changes to negative continuously as the chemical potential gets
larger than 0.42~GeV. These behaviors manifest that the chiral
symmetry breaking phase transits to the chiral symmetry preserving
phase at chemical potential $\mu=0.24$~GeV, and the phase transition
is of first order, and that the diquark channel of Wigner-vacuum is
stable while $\mu<0.42$~GeV, and the phase transition is of second
order. Therefore we can reach a conclusion for the massless-QCD
phase transition in the present truncated D-S equation approach as:
there is a first order phase-transition of chiral restoration at a
quite large chemical potential and color superconductivity phase
emerges as a second order phase transition at a very large chemical
potential where chiral symmetry has already been restored.

\begin{figure}[!htb]
\centering
\includegraphics[scale=1.0]{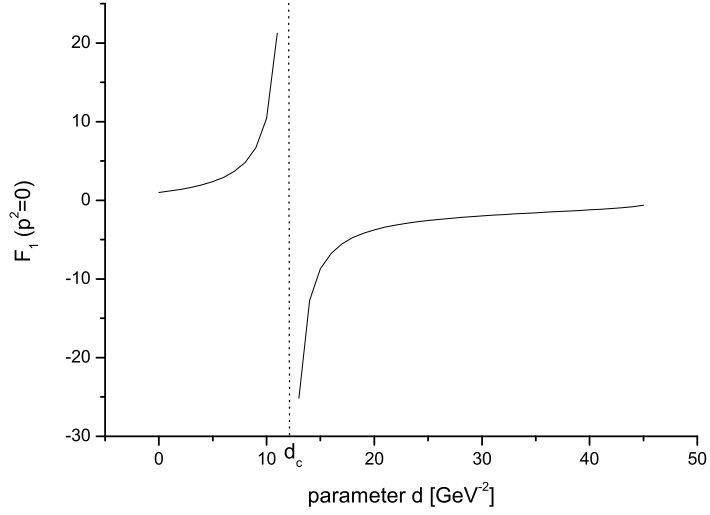}
\caption{\label{5} Coupling constant $d$ dependence of the
susceptibility $F_{1}$ at zero chemical potential and zero momentum}
\end{figure}

\begin{figure}[!htb]
\centering
\includegraphics[scale=1.0]{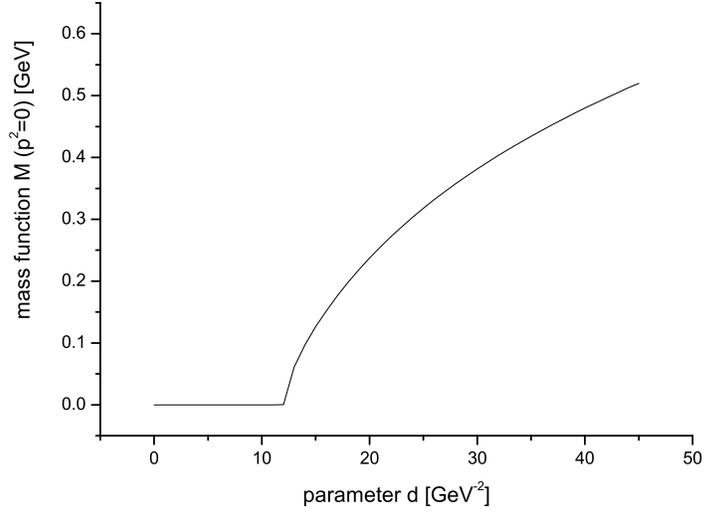}
\caption{\label{6} Coupling constant $d$ dependence of the dynamical
mass function at zero chemical potential and zero momentum}
\end{figure}

To make our criterion more robust, we also investigate the coupling
constant dependence of the chiral susceptibility $F_{1}$ of Wigner
solution at $\mu=0$ and that of the Nambu dynamical mass function
$M=\Delta_{1N}/A_{N}$ at $\mu = 0$. The obtained results are
displayed in Fig.~5, Fig.~6, respectively. It is evident that if the
interaction strength parameter $d < 12.2~{\mbox{GeV}}^{-2}$, the
chiral symmetry preserving phase (with dynamical mass $M=0$) is
stable. Only if the interaction in the infrared region is strong
enough (in the present numerical case with $\omega = 0.4$~GeV, the
$d$ should be larger than $12.2~{\mbox{GeV}^{-2}}$), can the chiral
symmetry breaking take place. Meanwhile, it should be emphasized
that while the critical coupling constant shown in Fig.5 can be
easily found by straightforward solving the D-S equation for
dynamical mass function (see Fig.6), the critical chemical potential
shown in Fig.3 can not be obtained by solving the D-S equation for
dynamical mass function (see Fig.2).

In summary, by solving the Dyson-Schwinger equations, we have
studied the chemical potential dependence of the solutions of the
D-S equation and of the Wigner-vacuum susceptibility in the chiral
and diquark channels as well as the effect of the interaction
strength. It shows that if the chemical potential of the system is
not very large and the interaction in the infrared region is strong
enough, the QCD vacuum is in the chiral symmetry breaking phase. If
the chemical potential gets larger and reaches a critical value, the
chiral symmetry can be restored by means of a first order phase
transition. If the chemical potential is much larger so as to arrive
at another critical value, the color superconductivity phase
emerges. Meanwhile the process of the chiral symmetry breaking is
proposed to be a positive feedback with respect to the perturbation.
Considering the criterions to characterize the chiral phase
transition, we would like to mention that it is not advisable to
introduce any approximate expression for free energy to judge which
solution (Wigner solution or Nambu solution) is free energy
favorable. First, no body knows how to construct an exact expression
for free energy. It is more serious that an approximate expression
can hardly satisfy the fact that Wigner-vacuum should always belong
to the ground state Hillbert-subspace no matter the true vacuum is
degenerate or not. Second, more importantly, it is not necessary to
construct a free energy expression for understanding phase
transition because the D-S equations (not solutions of D-S
equations) contain naturally the full phase information of the
dynamical system. For instance, as we have shown in this paper, the
susceptibilities of the Wigner-vacuum can be widely used to
understand dynamical symmetry breaking of the physical vacuum.

\bigskip

%\begin{acknowledgments}
This work was supported by the National Natural Science Foundation
of China (NSFC) under contract Nos. 10425521 and 10135030, the
Major State Basic Research Development Program under contract No.
G2000077400, the Key Grant Project of Chinese Ministry of
Education (CMOE) under contact No. 305001 and the Research Fund
for the Doctoral Program of Higher Education of China under grant
No. 20040001010. One of the authors (YXL) thanks the support of
the Foundation for University Key Teacher by the CMOE, too. The
authors are also indebted to Mr. Lei Chang for his helpful
discussions.

%\end{acknowledgments}

\end{document}